# Raman studies on nanocomposite silicon carbonitride thin film deposited by RF magnetron sputtering at different substrate temperatures

**Arnab Sankar Bhattacharyya, Suman Kumari Mishra**[*]

National Metallurgical Laboratory, Jamshedpur- 831007

**Abstract**

Raman studies of nanocomposite SiCN thin film by sputtering showed that with increase of substrate temperature from room temperature to 500°C, a transition from mostly $sp^2$ graphitic phase to $sp^3$ carbon took place which was observed from the variation of $I_D/I_G$ ratio and the peak shifts. This process resulted in the growth of $C_3N_4$ and $Si_3N_4$ crystallites in the amorphous matrix which led to increase in hardness and modulus obtained through nanoindentation. However, at a further higher temperature of 600°C, again an increase of $sp^2$ C concentration in the film was observed and the H and E values showed a decrease due to increased growth of graphitic carbon phase. The whole process got reflected in a modified four stage Ferrari Robertson model of Raman spectroscopy.

## 1. Introduction

Although carbon can be classified mainly into two allotropic phases -graphite like $sp^2$ hybridized carbon and diamond like $sp^3$ hybridized carbon, a mixed state of $sp^2$ and $sp^3$ hybridized carbon is also possible. The clustering of graphite leading to decrease in grain size is also a significant phenomenon leading to the polymorphs of carbon. Secondly, carbon in the atmosphere reacts with hydrogen leading to a formation of hydrocarbon (CH). All these phenomena led to classify carbon into number of phases [1, 2].

The shape of Raman spectrum is considered to be dependent on the clustering of the $sp^2$ phase, bond disorder, presence of $sp^2$ rings or chains and $sp^2/sp^3$ ratio. The Raman spectrum of the Si-C-N films contained mainly a G (graphitic) and D (disorder) peak which are also found in amorphous carbon films. The origin of the peaks is related to the total optic modes of graphite which is a sum of different modes as given below in eq 1.

$$\Gamma = A_{2u} + 2B_{2g} + E_{1u} + 2E_{2g} \tag{1}$$

$E_{2g}$ modes are Raman active and are observed at 42 and 1581 cm$^{-1}$. $A_{2u}$ and $E_{1u}$ are IR active and observed at 867 and 1588 cm$^{-1}$. $B_{2g}$ are optically inactive. E-symmetry modes have in plane atomic displacement while A and B have out of plane displacements. The G peak is associated with $E_{2g}$ zone centre stretching modes of graphite while the D (disorder) peak is due to $A_{1g}$ breathing modes which may be induced by sp$^3$-bonded carbon substitutional nitrogen atom and limitations in crystallite size. This D peak only occurs in graphites with small crystallites. This is due to the fact that disorder-induced mode corresponds to a peak in the vibrational density of states (DOS) of graphite and is observed when the crystallite size is sufficiently small so that the selection rules are relaxed to allow phonons with non-zero wave vectors to contribute to the Raman spectrum[1-5].

Si-C-N nanocomposite coatings have showed improved properties like high thermal conductivity, thermal stability (upto 1500°C), oxidation stability, high hardness, wide band gap chemical inertness, promising wetting behaviour, wear resistance, excellent chemical stability, and promising mechanical and thermal and optical properties[5,6]. The Si-C-N system is expected to have different superhard phases namely SiC, β-Si$_3$N$_4$ and β-C$_3$N$_4$ phase. The β-C$_3$N$_4$ and β-Si$_3$N$_4$ having the similar crystal structures are expected to be miscible in each other and form Si-C-N phase with excellent hardness [7, 8]. The reason for these unique properties shown by Si-C-N is complex covalent bonding, cross-linking of Si, C and N atoms, low oxygen diffusion, small grain size, large volume fraction of the interface, compressive residual stress, inter grain amorphous layer, growth orientation and solid solubility of different phases[5-8]. Si-C-N coatings are desired for many industrial applications such as MEMS, turbine engines, blades and wear resistant coatings for automotive industry to enhance the life as well as the performances of the components. They also find application in MEMS device fabrication in the form of field emission displays, catalyst support high temperature semiconducting device, metal and polymer matrix composites and high temperature applications [9, 10]. Both crystalline and amorphous or nanostructured Si-C-N compounds have been prepared. Thin film depositions of Si-C-N have been carried out by Plasma and ion assisted deposition. Chemical Vapour Deposition, magnetron sputtering, microwave and electron cyclotron resonance PECVD, ion implantation and pulsed laser deposition and Rapid thermal Chemical vapour deposition. Nanocrystalline $Si_2CN_4$ has been prepared by metal organic CVD [11-17].

The Raman studies are effective tools for structural analysis of particularly amorphous and nanocrystalline films, where X-Ray and other diffraction technique by and large show no peaks. Raman spectroscopy has been used earlier to characterize SiCN both in bulk and thin film form. Raman analysis show that Si-N, C=N, nitrogen-bound $sp^{2-}$ and $sp^{3-}$ hybridized carbon bonds were formed in the films. The free carbon phase forming in the polycermer derived bulk SiCN has been studied with the help of Raman spectroscopy. The effects of pyrolysis temperature and $CH_4/N_2$ gas mixture ratios have also been studied and reported [18-26].

The present communication discusses the Raman studies of SiCN thin films deposited at different substrate temperatures. At different substrate temperatures, the phase, bonds of the deposited film are different and they play a direct role in the physical, mechanical and microstructure behavior of the films. A correlation and explanation of the Raman spectra observations have been done Ferrari Robertson model and mechanical behavior by nanoindentation and microstructure has been carried out.

**Experimental**

Si-C-N Coatings were deposited on Si (100) substrate by RF magnetron sputtering (HHV, Bangalore, India) using single 500 mm sintered SiC target under argon and nitrogen plasma. For deposition the vacuum chamber was first evacuated to $1 \times 10^{-4}$ Pa pressure. Thin films were deposited on Si (100) substrate at 1 Pa pressures in $Ar/N_2$ atmosphere in 1:9 ratios. The temperature was varied from room temperature to 600°C at 400 watt RF power. The deposition was done for three hours. Nanoindentation tests were carried out in continuous stiffness mode using XP Nanoindenter MTS, USA. Nanoindentation experiments were done at varying depth of the film using Berkovich indentor with strain rate of 0.05 $s^{-1}$. Poisson's ratio was taken as 0.25 for calculations of all the parameters. A harmonic displacement of 2nm at a frequency of 45 Hz was initialized for the experiments. Tip calibration and microscopic calibrations for contact area and positioning respectively were carried out after every 10 indents. The hardness and modulus was calculated from load depth curve using Oliver and Pharr method.

Microstructural investigation was done by Transmission Electron Microscope (TEM) by Phillips (EM200). The TEM was carried out at 200 kV. For TEM studies Si-C-N deposition was also done simultaneously with Si substrate on the carbon coated copper grids at different substrate temperatures such as RT, 300°C, 500°C and 600°C respectively. Raman spectroscopy (Nicholet XR) using a He-Ne laser of 632.8 nm was used for structural characterization the Si-C-N coatings. A 25μm pin hole with an Olympus microscope was present to locate the area of interest.

## 3. Results and discussions

The Raman spectra of Si-C-N coatings at different substrate temperatures along with their deconvolution in the range of 1000-1750 cm$^{-1}$ using 2 Gaussian fits are given in Fig1 & Fig2. The G and D peak position and the integral intensity ratio ($I_D/I_G$) of the two peaks are given in Table 1 along with the substrate temperatures and nanoindentation hardness and modulus of the coatings. Fig 3 shows the variation of G, D peak position and the $I_D/I_G$ ratio variations with substrate temperatures

A broad peak at 2900cm$^{-1}$ was observed in all the spectra which has been reported to be arising is due to C-H bond due to residual water vapour trapped in the coatings. However some have called this peak as the second order D and G peak [27].

For a detailed analysis of the spectra, we first have to look into the three stage model proposed by Ferrari and Robertson for carbon. The transition from one carbon form to another is due to the variation in sp$^2$/sp$^3$ carbon percentage and results in shift in G peak position, change in $I_D/I_G$ ratio and crystallite size. The three stage model proposed by Ferrari and Robertson explains the transition taking place during this allotropic change associated with carbon [1].

The first stage is due to the transition from graphitic phase to nc-graphitic phase. A shift of the G band position to higher wavenumbers and appearance of D peak occurs accompanying this change. The $I_D/I_G$ ratio at this stage is related to the crystallite size (La) according to the Tuinstra and Koenig [28] equation given by

$$\frac{I_D}{I_G} \propto \frac{1}{L_a} \qquad (2)$$

The second stage consists of nc-graphite to a-C transition which is accompanied by a shift of the G peak to lower wavenumbers and decrease of $I_D/I_G$. As the *D* peak arises from aromatic rings, starting from graphite, $I_D/I_G$ will increase with increasing disorder, according to TK. For more disorder, clusters decrease in number become smaller and more distorted. As the *G* peak is just related to the relative motion of C $sp^2$ atoms, the $I_D$ will therefore decrease with respect to $I_G$ and the TK relationship will no longer hold. For small $L_a$, the *D*-mode strength is proportional to the probability of finding a sixfold ring in the cluster, that is, proportional to the cluster area and the equation proposed by Ferrari and Robertson (FR) [1] as given in eq (3) is applicable at this stage

$$\frac{I_D}{I_G} \propto L_a^2 \qquad (3)$$

The last stage consists of transition from a-C (amorphous carbon) to ta-C (tetrahedral-amorphous carbon) leading to an increase in G peak position from 1510cm$^{-1}$ to 1570cm$^{-1}$. This shift to higher wavenumber is due to the change of sp$^2$ configuration from rings to olefinic group [1-4]. The $I_D/I_G$ very low in this case and continues to follow the FR equation. This model has been used to explain the spectra arising in case of Si-C-N coatings as given below. The reason for using the existing model for Si-C-N was the parameters like clustering of the *sp$^2$* phase; bond disorder, presence of $sp_2$ rings or chains, and the $sp^2/sp^3$ ratio on which the Raman spectrum of Carbon is considered to depend are also the main contributions for the variation in the structural properties of Si-C-N. Moreover, the phenomenon occurring with the increase in substrate temperature is very similar to the *amorphization trajectory* ranging from graphite to *ta*-C (or diamond) in the three stages of the FR model. The relative concentration of Si and N in the SiCN films and their impact on the bonding behaviour and the on the Raman spectra has been discussed in our earlier publication where SiCN films deposited in RF and DC modes were compared [29].

In the spectra shown in Fig 1(a) corresponding to RT deposition, a G band position at 1571cm$^{-1}$ and D band position at 1350 cm$^{-1}$ was observed. An $I_D/I_G$ ratio of 2.14 was obtained (Fig 2(a)) which corresponds to mainly nanocrystalline (nc) graphitic phase and low percentage of sp$^3$ hybridized carbon falling in the stage-I of the model.

The $I_D/I_G$ ratio follows the TK equation in this region, the increase of $I_D/I_G$ ratio with substrate temperature from RT (Fig 3) indicated the lowering of the graphitic crystallite size due to higher nitrogen incorporation. The D band which is a parameter for disorder was also found to increase with substrate temperature in this stage due to carbon segregation as disordered or nano-crystalline graphitic clusters.

On reaching to 300$^o$C, the stage–II of the model starts where crystallite size continues to lower further which results in disappearance of nc-graphitic phase and formation of a-SiCN (amorphous) and the $I_D/I_G$ ratio follows the FR equation in this region. The $I_D/I_G$ ratio therefore shows a decrease to a value of 1.78 at 300$^o$C (Fig 3). A more detailed explanation for the TK→ FR is as follows: The *D* peak arises from aromatic rings. Starting from graphite, $I_D/I_G$ will increase with increasing disorder, according to TK. For more disorder, clusters decrease in number become smaller and more distorted. As the *G* peak is just related to the relative motion of C *sp$^2$* atoms, the *I*(*D*) will now decrease with respect to *I*(*G*) and the TK relationship will no longer hold. For small *La*, the *D*-mode strength is proportional to the probability of finding a six-fold ring in the cluster, that is, proportional to the cluster area and hence FR equation becomes valid. The variation of crystallite size with substrate temperature has been explained in detailed with the help of TEM studies in our previous publication [30].

The prominent separation between the D and G positions which is the evidence of graphitization gradually decreases and almost a single band formation at 1500 cm$^{-1}$ occurs at 500$^o$C (Fig 1).This corresponds to the end of stage-II and start of stage-III where the formation sp$^3$ hybridised tetrahedral amorphous carbon (*ta*-C) in the form of nucleation and growth of β-C$_3$N$_4$ crystallites in the a-SiCN matrix starts. This result in decrease in $I_D/I_G$ ratio but a shift of the G band position to higher wavenumbers.

At 600$^o$C however, a further increase in the G band position also accompanying an increase in $I_D/I_G$ ratio to 1.42 is observed (Fig 3). This is due to the formation and growth of crystalline and nc-graphitic phase due to escape of nitrogen from the films. The D peak position shows a decrease as the concentration of more stable phase of $sp^2$ carbon increases in comparison to $sp^3$ carbon. This corresponds to an additional stage-IV of the model (Fig 3).

The variation of $I_D/I_G$ ratio, D peak and G peak position with substrate temperature indicating the four stages of the model for SiCN coatings is similar to the three stage model proposed by Ferrari and Robertson. An explanation of the peak shifts and variation of $I_D/I_G$ ratio is given in the next section in terms of the structural changes occurring causing an effect in the polarization as discussed.

The $sp^3$ hybridised carbon has got tetrahedral symmetry consisting of single bonds and hence the possibility of movement is more compared to $sp^2$ hybridised carbon where the motion is restricted due to double bonds consisting of a planar symmetry as found in graphite which justifies for the fact that $sp^3$ carbon has a higher degree of freedom compared to $sp^2$ carbon

Deposition at higher temperature results in nitrogen to react the graphitic carbon initially present and formation of $sp^3$ hybridised carbon in the form of $\beta$-$C_3N_4$ as explained in the earlier section. This $sp^2$ to $sp^3$ transition initiates the doubly bonded carbon atom C=C at one of the corners of the hexagonal graphitic structure to transform into to a tetrahedral phase. This leads to an overall increase of the degrees of freedom of the system. Thus the polarization ellipsoid also undergoes a change in shape leading to a distortion in the stretching and breathing modes. The higher will be the conversion of the graphitic phase to amorphous phase the higher is the distortion causing the D and G peaks come closer and forming a band centered around 1500 cm$^{-1}$ corresponding to the average energy of the D and G bands. This process continues till the stage III is reached where the percentage of the $sp^3$ content in the film is high enough to make polarization ellipsoid start regaining its shape according to the tetrahedral symmetry of the $sp^3$ carbon. This will give the $sp^3$ bonded carbon an opportunity to undergo stretching causing an increase in energy and a shift of the G band to higher wavenumber [31].

TEM studies of the Si-C-N coatings exhibited the presence of a combination of amorphous microstructure and crystalline features. The microstructural morphology and phases underwent interesting changes on variation of substrate temperatures. Deposition of mostly graphitic and nanocrystalline graphitic phase occurred at room temperature (Fig 4 (a)). The inset SAED pattern showed polycrystalline rings due to a mixed phase of SiC and $sp^2$ hybridized carbon (graphite). Single crystal spot pattern due to large graphitic crystals were also observed.

Depositions at higher substrate temperatures led to disappearance of large graphitic crystallites which were found at RT and a transition to amorphous carbon started. The amorphous nature was higher due to higher nitrogen incorporation at higher temperature which led to finer crystallites. The a-C (amorphous carbon) reacted with Si and N to form an a-SiCN matrix. Nucleation and growth of $\beta$-$Si_3N_4$ nanocrytallites which initially started to take place at RT was more at higher temperatures. Evidence of formation of nanocrystalline $\beta$-$C_3N_4$ phases was found which gave a polycrystalline ring pattern shows the TEM micrographs and corresponding SAED patterns of coatings deposited at a substrate temperature of 500$^o$C (Fig 4(b)). Increasing the temperature further to 600$^o$C showed the presence of $Si_3N_4$ and $C_3N_4$ with larger sizes compared to 500$^o$C, as expected. However, some nitrogen desorption from the film occurred at 600$^o$C as $C_3N_4$ started losing nitrogen at higher temperature which led to deposition of graphitic phase (Fig 4(c)). The TEM observation thus supported the Raman spectra findings above which showed the transition from mostly $sp^2$ graphitic phase to sp3 carbon forming carbon nitride from room temperature to 500$^o$C and then again increase in $sp^2$ C A study involving XPS and AFM to correlate the changes taking place in the structural and mechanical properties is published elsewhere [32, 33].

Fig 5 (a) and (b)) gives the nanohardness (H) and modulus (E) values of coatings deposited on silicon substrates at various deposition temperatures in RF mode. The values given are the average of four indentations. The pressure and power were kept constant at $1\times10^{-2}$ mbar and 400 W respectively. H and E were found to increase with the substrate temperature from RT showing a hardness of 10GPa and modulus 100GPa to a highest hardness of 21GPa and 240GPa. On increasing the temperature further to 600$^o$C however showed a decrease in the H (8GPa) and E (105GPa) values. The higher value of hardness at 500$^o$C compared to room temperature and 600$^o$C is attributed to small particle size in the film at 500$^o$C substrate temperature and the presence of Si-N and C-N phases, whereas the lower hardness and modulus at room temperature and 600$^o$C was due to presence of soft graphite phase in the SiCN matrix of the film, as observed through TEM and Raman studies. A detailed nanoindentation studies of SiCN films deposited on different substrates is published elsewhere [30].

## 4. Conclusions

The three stage Ferrari-Robertson model applicable for carbon has been used in the case of SiCN coatings deposited at different substrate temperatures by magnetron sputtering. Interestingly after following the three stages of amorphization trajectory a fourth stage is also observed due to recrystallization of graphitic phase. The phenomenon occurring was confirmed by TEM studies and a change in hardness and modulus determined from nanoindentation studies were observed.

**Acknowledgement**

The author wishes to thank Council of Scientific and Industrial Research, India for financial support.

**Legends to figures**

Fig 1 Raman spectra of SiCN coating deposited on Silicon substrates deposited at (a) RT (b) 300$^{o}$C, (c) 500$^{o}$C and (d) 600$^{o}$C substrate temperatures.

Fig 2: Two Gaussian fit deconvolution of Raman spectra of SiCN coating deposited on Silicon substrates deposited at (a) RT (b) 300°C (c) 500 °C and (d) 600°C substrate temperatures

Fig 3: Proposed four stage model for substrate temperature variation in Si-C-N coating

Fig 4: Transmission Electron Micrographs of Si-CN coatings at (a)Room Temperature (b)500°C and (c)600°C substrate temperatures

Fig 5: (a) Hardness and (b) Elastic Modulus profiles obtained through nanoindentation of Si-C-N coatings deposited at different substrate temperatures

| s.no | Temperature (°C) | D-band (cm$^{-1}$) | $I_D/I_G$ (area ratio) | G-band (cm$^{-1}$) | H (GPa) | E (GPa) | stage |
|---|---|---|---|---|---|---|---|
| 1 | 30 | 1350 | 2.14 | 1571 | 10 | 100 | I |
| 2 | 300 | 1375 | 1.78 | 1575 | 14 | 140 | II |
| 3 | 500 | 1390 | 0.90 | 1522 | 21 | 240 | III |
| 4 | 600 | 1343 | 1.42 | 1570 | 8 | 105 | IV |

Table 1: Variation of Raman parameters with substrate temperature

**Figure-1**

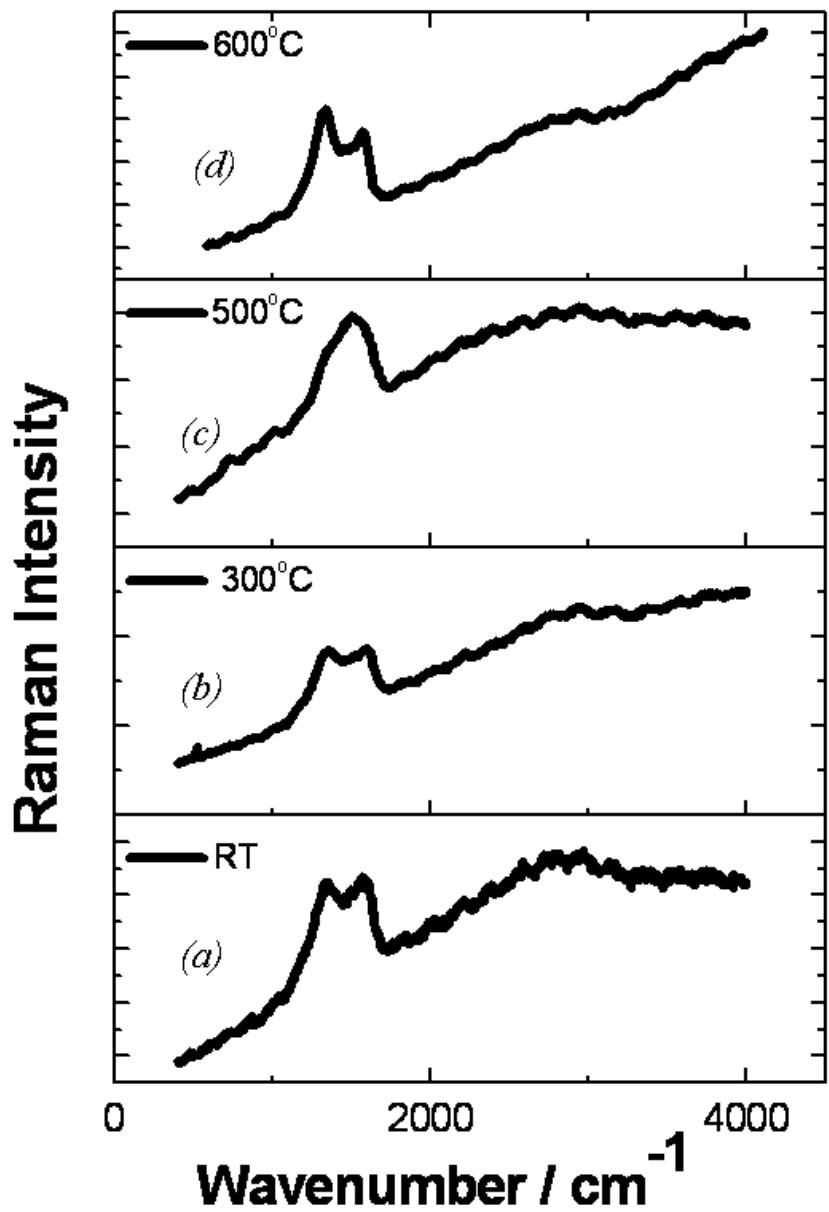

**Figure - 2**

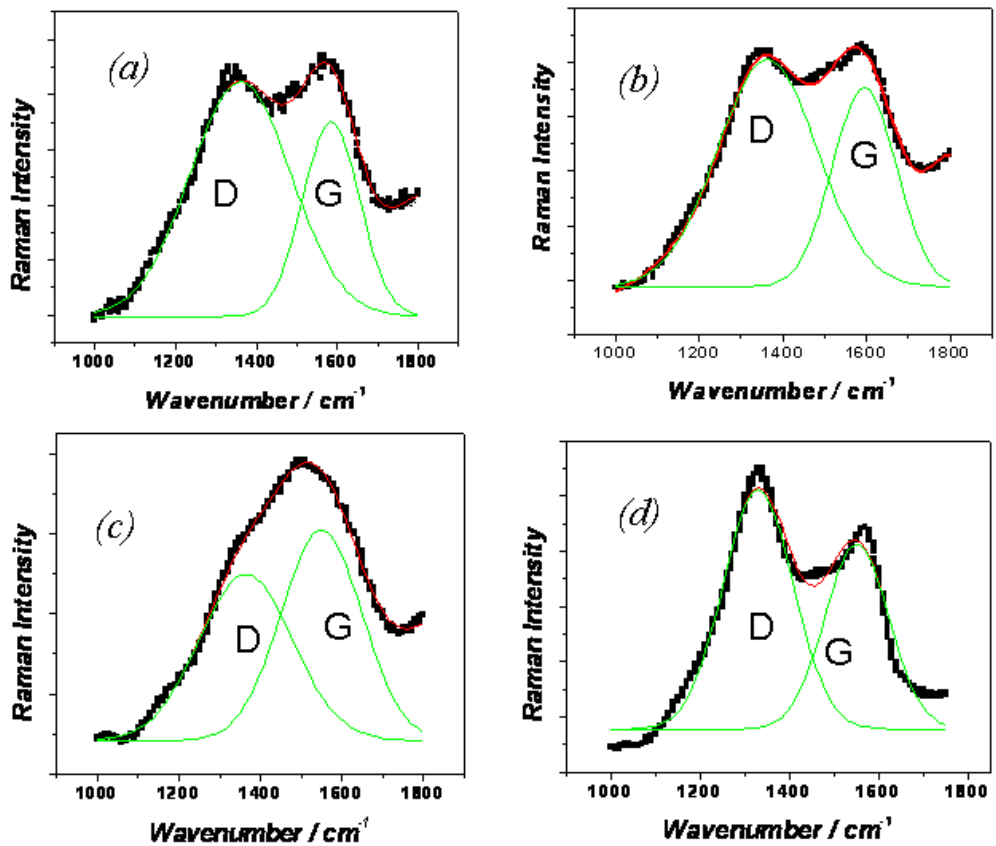

Figure - 3

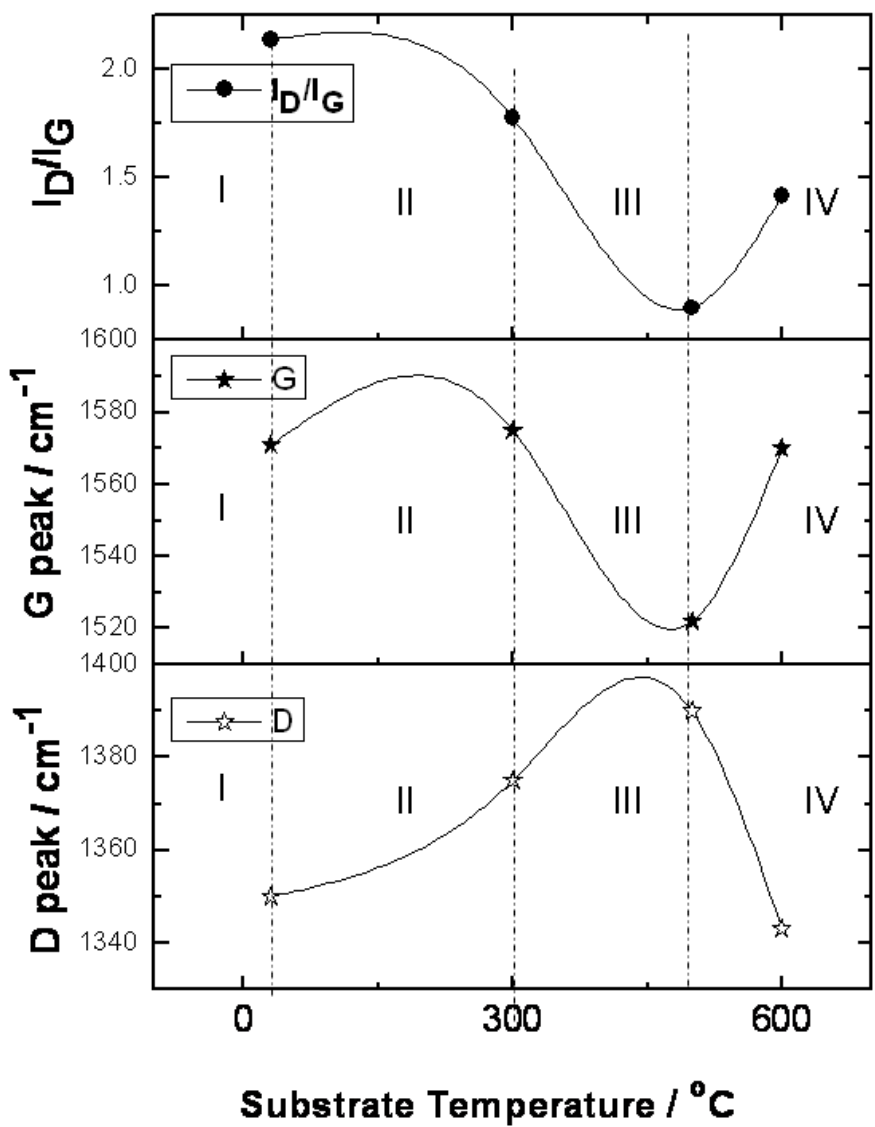

# Figure - 4

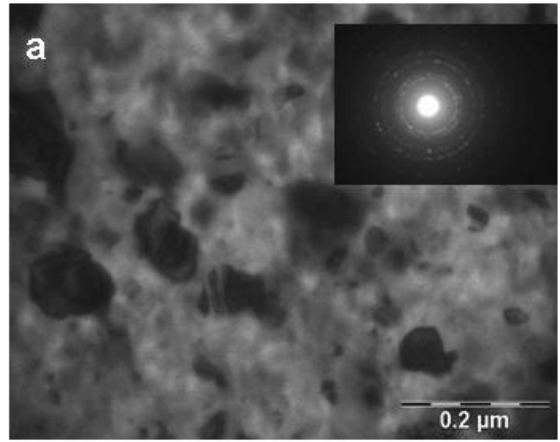

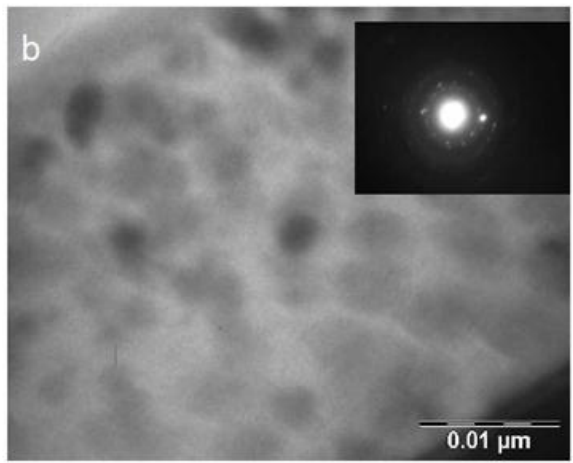

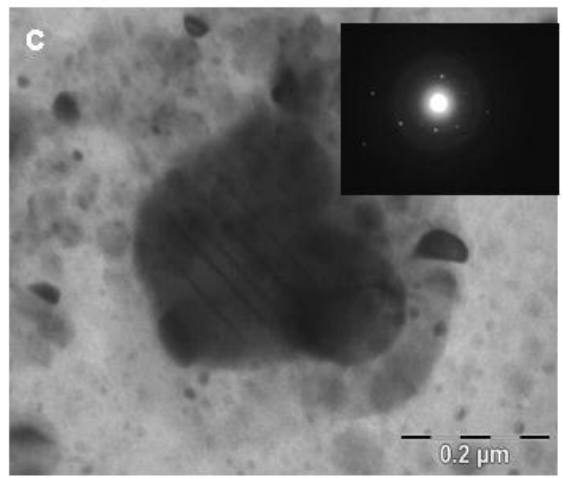

**Figure - 5**

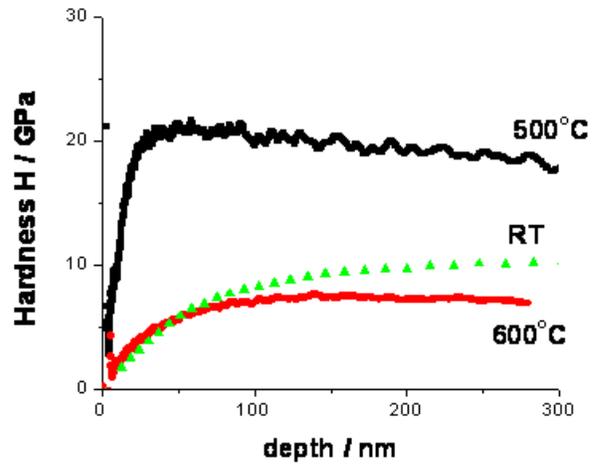

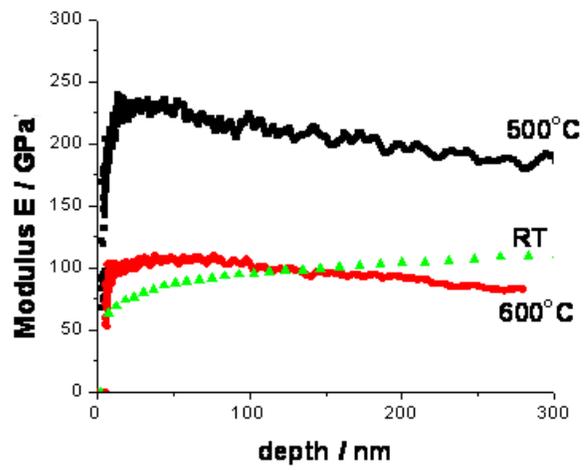